\documentclass[pre,aps,showpacs,twocolumn]{revtex4}
\usepackage{epsfig}
\usepackage{bm}

\begin{document}

\title{``Clusterization" and intermittency of temperature fluctuations in turbulent convection}

\author{\small A. Bershadskii$^{1,2}$, J.J. Niemela$^1$, A. Praskovsky$^3$ and K.R. Sreenivasan$^1$}
\affiliation{\small {\it $^1$International Center for Theoretical
Physics, Strada Costiera 11, I-34100 Trieste, Italy}\\
{\it $^2$ICAR, P.O. Box 31155, Jerusalem 91000, Israel} \\ {\it
$^3$National Center for Atmospheric Research, P.O. Box 3000,
Boulder, CO 80307, USA}}

\begin{abstract}
Temperature time traces are obtained in turbulent thermal
convection at high Rayleigh numbers. Measurements are made in the
midplane of the apparatus, near the sidewall but outside the
boundary layer. A telegraph approximation for temperature traces
is generated by setting the fluctuation amplitude to 1 or 0
depending on whether or not it exceeds the mean value. Unlike the
standard diagnostics of intermittency, the telegraph approximation
allows one to distinguish the tendency of events to cluster
(``clusterization") from their large-scale variability in
amplitude. A qualitative conclusion is that amplitude
intermittency might mitigate clusterization effects.

\end{abstract}


\maketitle

\section{Introduction}
We consider turbulent convection in a confined container of
circular crosssection and 50 cm diameter. The aspect ratio
(diameter/height) is unity. The sidewalls are insulated and the
bottom wall is maintained at a constant temperature, which is
higher by a small amount $\Delta T$ than that of the top wall. The
working fluid is cryogenic helium gas. By controlling the
temperature difference between the bottom and top walls, as well
as the thermodynamic operating point on the phase plane of the
gas, the Rayleigh number (Ra) of the flow could be varied between
$10^7$ and $10^{15}$. We measure temperature fluctuations at
various Rayleigh numbers towards the upper end of this range, in
which the convective motion is turbulent. Time traces of
fluctuations are obtained at a distance of 4.4 cm from the
sidewall on the center plane of the apparatus. This position is
outside of the boundary layer region for the Rayleigh numbers
considered here. More details of the experimental conditions and
measurement procedure can be found in Ref. \cite{NS1}.

A significant part of convection, even at the high Rayleigh
numbers that concern us here, is due to plumes \cite{LK}. We use
the term here merely to denote an organized activity of convection
without implying much about their three-dimensional shapes and
sizes, or the parameters on which they scale, though a few
comments will be made momentarily. The primary goal of the paper
is to learn about the tendency of the plumes to cluster together
(``clusterization").

The upper part of Fig.\ 1 shows a short segment of temperature
fluctuations at $Ra = 1.5 \times 10^{11}$. There are four
large-scale events within this segment (marked by the letters
A-D), and we imagine them to be the manifestation of large-scale
plumes. Each event consists of several subevents, and there also
exist a number of small events marked a-i. While it may well be
that the subevents deserve to be considered separately, we regard
them collectively here. Under these circumstances, it is clear
that a typical life-time of the large events is of the order of 8
seconds. Noting from Ref. \cite{NS2} that the mean speed of the
large-scale circulation (``mean wind") for these conditions is
about 6 cm/s, a typical length scale of these events is of the
order of 50 cm, which is the characteristic dimension of the
apparatus. That is, if these large-scale plumes originate from the
boundary layer, it is as if the entire boundary layer on the
bottom wall participates once in a while in this activity that we
have called the large-scale plume. The maximum temperature in
these large-scale plumes is a fraction of the excess temperature
of the bottom plate (namely $\Delta T/2$), so, presumably, the
fluid that is participating in the formation of a typical
large-scale plume comes from the top parts of the boundary layer;
or, if a plume does indeed come from the very bottom parts of the
boundary layer, it is already partially mixed by the time it
reaches the probe midway between the top and bottom walls. They
are certainly not small-scale events that scale on the thickness
of the boundary layer. This description does not apply to
small-scale plumes a-i, though, presumably, they too belong to the
same family. In Fig.\ 1, we show the case of hot plumes, that is,
the case when the wind at the measurement point arrives from the
hotter bottom plate. One can imagine that the wind direction could
be just the opposite, leading to the arrival at the probe of cold
plumes coming from the colder top plate. We have analyzed such
instances as well. Further, beyond a certain Rayleigh number, as
described in Ref. \cite{NS2}, the mean wind reverses itself
randomly, so that a probe permanently held at one position sees
hot plumes for some period of time and cold plumes for some other
period of time. We have analyzed the two parts separately by
stringing together only hot parts or the cold parts of a measured
temperature trace. The data for the two cases have been examined
separately. In each case, the telegraph approximation is related
to specific properties of the underlying physical processes
associated with hot or cold plumes, as they encounter the probe
during their motion.
\begin{figure}
\centering \epsfig{width=.45\textwidth,file=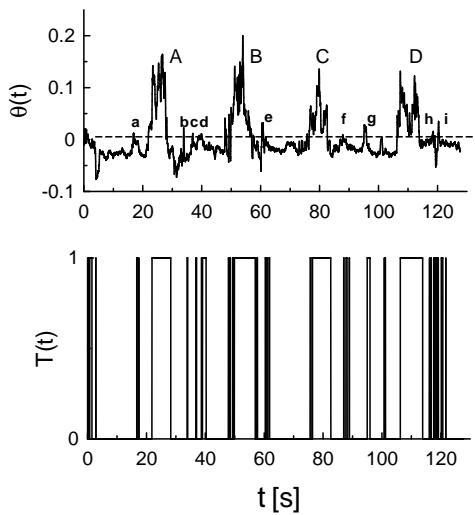}
\vspace{-3.5cm} \caption{An example of the measured temperature
time trace (upper part) and its random telegraph approximation
(lower part). In addition to the small plumes marked a-i, there
are several others such structures in the signals. They have not
been marked merely because they occur below the zero line.}
\end{figure}
\section{Random telegraph approximation}

A more detailed discussion of the plumes will be presented
elsewhere but we limit ourselves here to a discussion of their
tendency to cluster together occasionally. This is not obvious
from the piece of the temperature trace shown in Fig.\ 1, and a
longer trace crowds the plumes too much. It may be surmised that
the clusterization is indeed responsible for the mean wind in the
apparatus. In the usual methods of analysis of turbulent signals
\cite{my}, it is difficult to separate the clusterization effect
from the usual intermittency effects arising from amplitude
variability. To separate the two effects, we ignore the variation
of the amplitude from one plume to another and replace the
temperature trace of the type shown in the upper part of Fig.\ 1
by its random telegraph approximation, shown in the lower part.
This approximation is generated from the measured temperature by
setting the {\it fluctuation} magnitudes to 1 or 0 depending on
whether the actual magnitude exceeds the mean value (marked as
zero and shown by the dashed line in the upper part of Fig.\ 1).
Formally, for the temperature fluctuation $\Theta (t)$ (with zero
mean), the telegraph approximation $T(t)$ is constructed as
$$
T(t) = \frac{1}{2} \left(\frac{\Theta (t)}{|\Theta (t)|} + 1
\right). \eqno{(1)}
$$
By definition, $T$ can assume either 1 and 0. The telegraph
approximation can be generated by setting different ``thresholds"
than the mean. It turns out that most properties examined here are
reasonably independent of the threshold; this comment will be made
also at other specific places in the paper.
\begin{figure}
\centering \epsfig{width=.45\textwidth,file=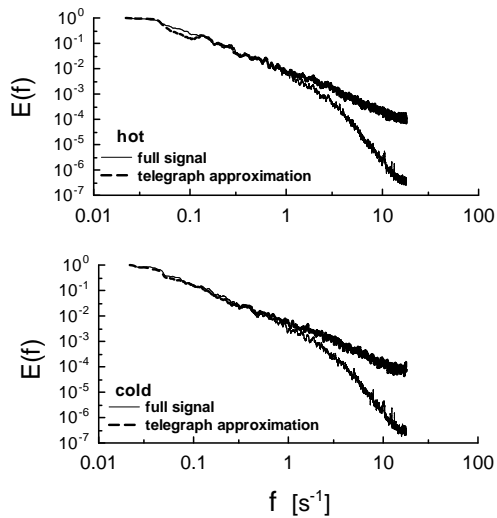}
\vspace{-3.5cm} \caption{Spectra of the telegraph approximation of
two realizations of the temperature trace compared with those of
the temperature trace itself. The ordinate has been shifted so
that the first points in both spectra coincide.}
\end{figure}
It may be useful to know how the conventional statistics for the
random telegraph approximation compare with those of the
temperature signal. Figure 2 compares the spectral densities,
$E(f)$, of the telegraph approximations with those of the full
signal. The comparisons are made separately for hot and cold
cases. It is clear that the spectra of the telegraph approximation
are close to those of the original signal in a significantly large
interval of scales. The main difference is that the telegraph
approximation is richer in spectral content above a certain
frequency. This is not difficult to understand from a visual
inspection of Fig.\ 1.

Of particular interest is the power-law behavior of the spectral
densities of the telegraph approximation. For both hot and cold
cases, they follow a power-law of the form
$$
E(f) \sim f^{-\beta}.             \eqno{(2)}
$$
Though this power-law behavior is clear from Fig.\ 2, we reproduce
in Fig.\ 3 the spectra of the telegraph signal, computed from
several records to attain better statistical convergence, in order
to emphasize the power-law scaling. The exponent $\beta = 1.38 \pm
0.02$. A reasonable shift of the threshold does not change the
spectral exponent $\beta$.
\begin{figure}
\centering \epsfig{width=.45\textwidth,file=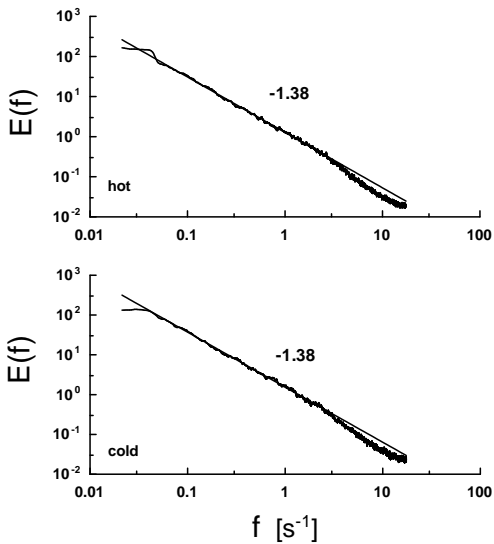}
\vspace{-3.5cm} \caption{Spectrum in the telegraph approximation
computed using twenty realizations of the temperature trace, for
the hot case (upper plot) and for the cold case (bottom part). The
straight lines (the best fit) show that the power-law
approximation (2) holds well for both hot and cold cases.}
\end{figure}
From the closeness of the spectra of the temperature trace with
its telegraph approximation, it is inferred easily that the former
has a spectral exponent of 1.38 as well, albeit over a smaller
range of scales. Observation of the temperature trace spectra with
such power law was first made in \cite{lib}, and was explored
theoretically in \cite{y}, and is now a well-known result.

The probability density function of the duration between events,
$\tau$, for the telegraph approximation is shown in Fig.\ 4. Data
for the hot and cold cases are given separately. The log-log scale
has been chosen to emphasize the power-law structure
$$
p(\tau) \sim \tau^{-\alpha}.   \eqno{(3)}
$$

For both hot and cold cases, we observe $\alpha = 1.37 \pm 0.03$.
A reasonable variation of the threshold in the vicinity of the
average value does not change the exponent $\alpha$.

It is known that, for non-intermittent cases (Ref. \cite{jen} and
Sec.\ 4), the relation between the exponents $\alpha$ and $\beta$
is given by
$$
\beta = 3-\alpha.    \eqno{(4)}
$$
If we substitute in (4) the value of $\alpha \simeq 1.37 $ (as
observed in Fig.\ 4) we obtain $\beta \simeq 1.63$. This is
considerably larger than that actual value measured in Fig.\ 3,
namely 1.38. This discrepancy is the object of interest to us
here; as already remarked, since there are no amplitudes involved,
it must be related to clusterization entirely (see also
\cite{jen}).
\begin{figure}
\centering \epsfig{width=.45\textwidth,file=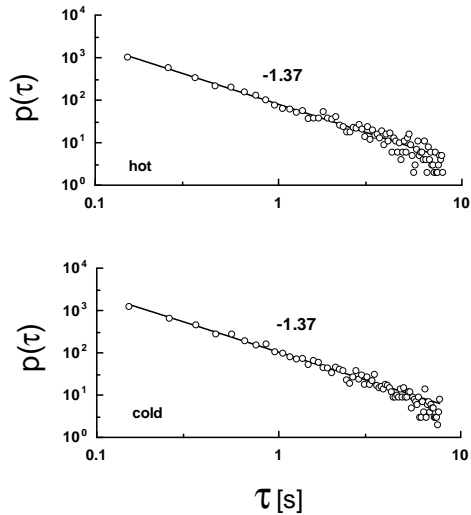}
\vspace{-3.5cm} \caption{The probability density function of the
duration $\tau$ for the telegraph approximation of the temperature
signal (for hot and cold cases). The straight lines (the best
fits) are drawn to indicate the scaling law (3).}
\end{figure}
It is of interest to note here that, for the telegraph
approximation of the temperature fluctuation in the turbulent
atmospheric boundary layer, we have about the same values of
$\alpha$ and $\beta$ as the present, while the spectral density of
the temperature trace has a roll-off rate of about 1.66
(consistent with the Kolmogorov-Obukhov-Onsager-Corrsin theory
\cite{my}). It is clear that the spectra of the temperature signal
and its telegraph approximation are closer in confined convection
than in atmospheric turbulence.

\section{Quantifying clusterization}

The difference between the observed telegraph spectral exponent
($\beta \simeq 1.38$) and its value given by Eq.\ (4) ($\simeq
1.63$) is a quantitative measure of clusterization \cite{jen} of
plume-like objects observed in temperature traces. This is a part
of intermittency.

Intermittency of the so-called temperature dissipation rate
\cite{my},\cite{sa} is characterized in turbulence by
$$
\chi = |\frac{dT^2}{dt}|.    \eqno{(5)}
$$
Following Obukhov \cite{my}, the local average
$$
\chi_{\tau} =\frac{1}{\tau}\int_{t}^{t+\tau} \chi (t) dt
$$
can be used to describe the intermittency of $\chi$. The scaling
of the moments,
$$
\frac{\langle \chi_{\tau}^q \rangle}{ \langle
\chi_{\tau}\rangle^q} \sim \tau^{-\mu_q},  \eqno{(6)}
$$
assuming that scaling exists, is a common tool for the description
of the intermittency \cite{my},\cite{sa}. Intermittent signals
possess a non-zero value of the exponent $\mu_q$. Of particular
interest is the exponent $\mu_2$ for the second-order moment.

The telegraph approximation is a composite of Heaviside step
functions, so the dissipation rate (5) is a composite of pulses
(i.e.\ delta-functions) located at the edges of the boxes of the
telegraph signal. For the uniform random distribution of the
pulses along the time axis, $\mu_q=0$. Non-zero values of $\mu_q$
mean that there is a clusterization of pulses. Figure 5 shows
dependence of the normalized dissipation rate $\langle
\chi_{\tau}^2 \rangle$ on $\tau$ for the telegraph approximation
of hot and cold signals. The straight lines (the best fits) are
drawn to indicate the scaling law (6), for $q=2$. It should be
noted that scaling interval for the dissipation rate is the same
as for the PDF and for the spectrum (cf.\ Figs.\ 3 and 4). Values
of the intermittency exponent $\mu_2$, calculated as slopes of the
straight lines in Fig.\ 5 is $\mu_2 \simeq 0.47 \pm 0.03$. The
relatively large value of the exponent $\mu_2$ suggests that the
clusterization of the pulses is quite strong.
\begin{figure}
\centering \epsfig{width=.45\textwidth,file=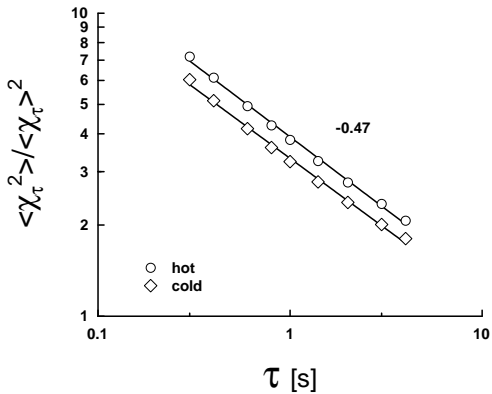}
\vspace{-5cm} \caption{Normalized second moment of the local
dissipation rate for the telegraph approximation plotted against
$\tau$, for cold and hot cases. The straight lines (the best fits)
are drawn to indicate the scaling law (6).}
\end{figure}
The temperature dissipation can be also characterized by the
``gradient" measure \cite{my},\cite{sa}
$$
\chi_r =\frac{\int_{v_r} (\bigtriangledown{T})^2 dv}{v_r},
$$
where $v_r$ is a subvolume with space-scale $r$ (for detailed
justification of this measure, see Ref.\ \cite{my}, p.\ 381 and
later). The scaling law of the moments of this measure are
important characteristics of the dissipation field \cite{sa}. By
Taylor's hypothesis \cite{my}, we can replace $dT/dx$ by
$dT/\langle u \rangle dt$ (where $\langle u \rangle$ is the mean
wind and $x$ is the coordinate along the direction of the wind),
and can define the dissipation rate as
$$
\chi_{\tau} \sim \frac{\int_0^{\tau} (\frac{dT}{dt})^2 dt}{\tau},
$$
where $\tau \simeq r/\langle u \rangle$. This, too, should follow
the scaling relation (6).

This definition has a problem in the telegraph approximation
because $dT/dt$ is composed of delta functions. Fortunately, one
is interested in the scaling of the discrete representation of the
dissipation field, given by
$$
\chi_{\tau} \sim \sum_{n=1}^\tau (\Theta_{n+1}-\Theta_n)^2 /\tau,
$$
where $n$ specifies an interval of space. This discrete definition
of $\chi_{\tau}$ avoids the problem with delta functions.
Obviously, for the telegraph signal, scaling exponents calculated
for the {\it pulse}-defined dissipation, Eqs.\ (5)-(6), and for
the discrete {\it spike-like} process are identical; this
observation is not true for the original temperature $\Theta (t)$.

\section{Clusterization and the spectrum}

The spectrum of the telegraph approximation can be related to the
probability distribution $p(\tau)$ through
$$
E(f) = \int W (t) G(ft) p(t) dt,   \eqno{(7)}
$$
where $G(ft)$ is a transform function and the weight function
$W(t)$ is supposed to have a scaling form
$$
W(t) \sim t^{\delta}.        \eqno{(8)}
$$
Since the quantities $G(f t)$ and $p(t)dt$ are dimensionless, one
can use dimensional considerations to find the exponent $\delta $
(cf.\ Ref. \cite{my}) through
$$
W(t) \sim \langle \chi \rangle \cdot t^2,   \eqno{(9)}
$$
and the use of (2),(3) and (7); this yields relation (4).

To estimate the clusterization correction on the relation between
scaling spectrum and $p(t)$, we should take into account two-point
correlations in the telegraph signal. This can be characterized by
the two-point correlation function $\xi (t)$. In a situation where
the two-point correlation function exhibits the scaling behavior
$$
\xi (t) \sim t^{-\gamma},   \eqno{(10)}
$$
the correlation exponent $\gamma$ is the same as the exponent
$\mu_2$. This is easily seen by the well-known result that the
correlation dimension \cite{proc} $D_2$ is related to $\gamma$
through
$$
D_2=1-\gamma,       \eqno{(11)}
$$
and to $\mu_2$ Ref. \cite{sa} via
$$
D_2 = 1- \mu_2,  \eqno{(12)}
$$
thus yielding $\gamma = \mu_2.$ To estimate the weight function
$W(t)$ with the same dimensional considerations as above, and to
take into account the two-point correlation (which characterizes
clusterization), we replace $\langle \chi \rangle$ by
$$
\langle \chi_t^2 \rangle^{1/2} \sim t^{-\mu_2/2}. \eqno{(13)}
$$
Replacing (9) by
$$
W(t) \sim \langle \chi_{t}^2 \rangle^{1/2} \cdot t^2, \eqno{(14)}
$$
we have
$$
W(t) \sim t^{2- \mu_2 /2}.              \eqno{(15)}
$$
The corresponding correction of Eq.\ (4) is
$$
\beta = ( 3-\mu_2/2) -\alpha.  \eqno{(16)}
$$
Using the value of $\mu_2 \simeq 0.47$ from Fig.\ 5 and the value
$\alpha \simeq 1.37$ from Fig.\ 4, we obtain
$$
E(f) \sim f^{-1.40},  \eqno{(17)}
$$
which compares well with the behavior found in Fig.\ 3 (see also
\cite{lib},\cite{y}).

\section{Summary and discussion}

The results discussed so far are for a fixed Rayleigh number. The
same situation seems to occur at other Rayleigh numbers. For those
Rayleigh numbers where there is a reversal of the wind, the
concatenated data corresponding to one given direction of the wind
follow the same statistics as well. Thus, the characteristics
discussed in the paper are generally valid for turbulent
temperature fluctuations at all Rayleigh numbers covered in the
measurements. The main conclusion is that the telegraph
approximation captures the main statistical features of the
temperature time trace obtained in convection. This approximation,
which gives a clear separation between clusterization and
magnitude intermittency, has been useful in demonstrating that
there is a significant tendency for the plumes to cluster
together. The telegraph approximation turns out to be useful here
because of the specific process of heat transport, which is
determined in large measure by the random motion of temperature
plumes. However, one can expect that this approximation (or its
modifications) may be also useful in the description of other
turbulent signals. The exponent $\mu_2$ for the telegraph
approximation, completely determined by clusterization, is about
0.47. This should be compared with the intermittency exponent
computed for the $\chi$-$\chi$ correlation of the full temperature
signal, which is about 0.36---consistent with similar estimates
available for passive scalars (see, e.g., Ref. \cite{pms}). The
clusterization exponent is thus larger than the classical
intermittency exponent. From this, one can infer that the
magnitude intermittency plays a smoothing role on the
clusterization effects within the scaling interval.\\

We thank L.J. Biven, J. Davoudi and V. Yakhot for useful
discussions.

\end{document}